\documentclass[runningheads]{llncs}
\usepackage{graphicx}

\usepackage{listings}
\newcommand\codefont{\ttfamily\fontsize{8}{9}\selectfont}

\lstdefinelanguage{Turtle}{
  comment=[f]{\#},
}

\usepackage{dirtytalk}
\usepackage{color,soul}
\usepackage{wrapfig}
\usepackage{csquotes}
\graphicspath{ {./images/} }

\begin{document}
\title{Event Notifications in Value-Adding Networks}
%
%
\author{Patrick Hochstenbach\inst{1,2}\orcidID{0000-0001-8390-6171} \and
Herbert Van de Sompel\inst{2,3}\orcidID{0000-0002-0715-6126} \and
Miel Vander Sande\inst{4,5}\orcidID{0000-0003-3744-0272} \and
Ruben Dedecker\inst{2}\orcidID{0000-0002-3257-3394} \and
Ruben Verborgh\inst{2}\orcidID{0000-0002-8596-222X}
}
\authorrunning{P. Hochstenbach et al.}
%
\institute{Ghent University Library, Ghent University, Ghent, Belgium \and
IDLab, ELIS, Ghent University - IMEC, Ghent, Belgium \and
DANS, Data Archiving and Networked Services, The Hague, The Netherlands \and
Meemoo, Vlaams Instituut voor het Archief, Ghent, Belgium \and
Dutch Digital Heritage Network, The Hague, The Netherlands
\email{\{Patrick.Hochstenbach,Herbert.VandeSompel,Ruben.Dedecker,Ruben.Verborgh\}@UGent.be}\\
\email{Miel.VanderSande@meemoo.be}}
\maketitle              
\begin{abstract}
Linkages between research outputs are crucial in the scholarly knowledge graph. They include online citations, but also links between versions that differ according to various dimensions and links to resources that were used to arrive at research results. In current scholarly communication systems this information is only made available post factum and is obtained via elaborate batch processing. In this paper we report on work aimed at making linkages available in real-time, in which an alternative, decentralised scholarly communication network is considered that consists of interacting data nodes that host artifacts and service nodes that add value to artifacts. The first result of this work, the “Event Notifications in Value-Adding Networks” specification, details interoperability requirements for the exchange of real-time life-cycle information pertaining to artifacts using Linked Data Notifications. In an experiment, we applied our specification to one particular use-case: distributing Scholix data-literature links to a network of Belgian institutional repositories by a national service node. The results of our experiment confirm the potential of our approach and provide a framework to create a network of interacting nodes implementing the core scholarly functions (registration, certification, awareness and archiving) in a decentralized and decoupled way.

\end{abstract}
\section{Introduction}
\label{Introduction}

In the 2004 paper \enquote{Rethinking Scholarly Communication} \cite{Rethinking}, Van de Sompel et al introduced the perspective of an open scholarly communication system in which decentralized service hubs add value to research outputs hosted in repositories by fulfilling the core functions of scholarly communication - registration, certification, awareness, archiving \cite{ROOSENDAAL} - in a decoupled manner and in ways they see fit. The fact that value of a certain kind was added was intended to be openly exposed and, as such, would  support creating bidirectional linkages between resources jointly involved in a life-cycle event, such as a preprint and a review thereof, an article and a copy of it in a long-term archive, etc. The authors understood that, in order for such an environment to be feasible, \enquote{the service hubs needed to be interconnected, as if they were part of a global scholarly communication workflow system}. 

A similar perspective is at the basis of the more recent \enquote{Linked Research on the Decentralized Web} \cite{CARVEN} and COAR's \enquote{Next Generation Repositories report} \cite{NextGenRep} but remains in sharp contrast with the status quo. The scholarly communication system does not provide real-time life-cycle information. To a large extent, linkages between related research outputs are revealed by means of post-factum, heuristic-bound mining processes across research output corpora (e.g., Cabernac et al efforts to discover preprint/publication links  \cite{paper_link_preprints_reviewed}), that are so laborious that only the happy few can consider providing them as production services.

\begin{wrapfigure}{r}{0.4\textwidth}
    \centering
    \includegraphics[width=0.4\textwidth]{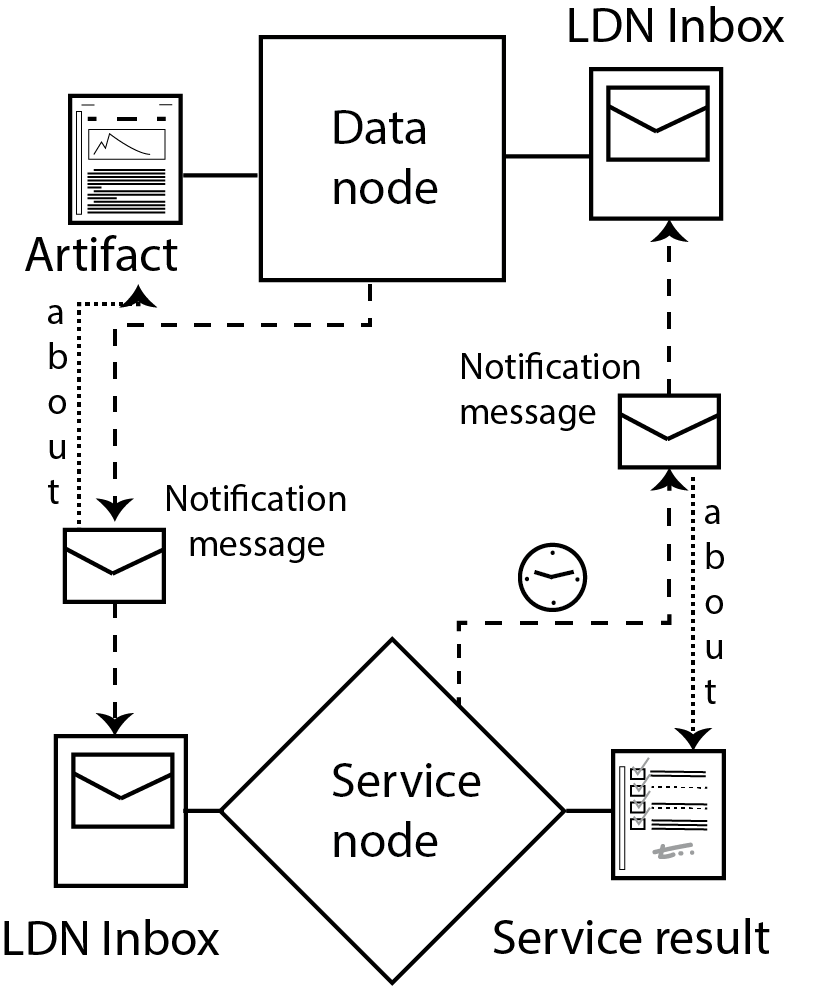}
    \caption{Data nodes and service node exchanging notification messages.}
    \label{fig:network}
\end{wrapfigure}

This paper introduces the \enquote{Event Notifications in Value-Adding Networks} \cite{MellonSpec} specification, which is a first step towards realizing the ability to expose life-cycle information for research outputs in real-time and hence to reveal linkages as they become known. In the network (see Figure \ref{fig:network}) considered by the specification, service nodes add value to artifacts hosted by data nodes and the nodes exchange messages with that regard: a data nodes asks a service node to add value to one of its artifacts, and the service node responds with information about the service result; a service node volunteers information about value it unilaterally added to an artifact hosted by a data node. For example, a researcher can request that a research output be peer-reviewed via a certification service or directly by selected peers; she can request that a search engine include the research output in its index; she can ask an archive to collect the output and store it for posterity; a service can inform the researcher that her research output was leveraged by another endeavor; a peer can inform the researcher about a new annotation made to the research output. In the scholarly realm, the nature of the added value crucially includes fulfilling the core functions of scholarly communication. But, while scholarly communication was the main driver for the specification, it is conceived in a generic way and allows for a wide range of added-value as well as applications in other domains such as cultural heritage. 

The messages between the autonomous systems that operate on the network are point-to-point, owing to its decentralized nature. The \enquote{Event Notifications in Value-Adding Networks} specification details interoperability regarding the messaging protocol and the communication payloads that are essential to allow the envisioned network to be functional. 

Our specification effort is generally motivated by the Decentralized Web \cite{socially_aware_cloud_storage} movement that emerged in reaction to the disconcerting evolution of the Web towards an environment in which a limited number of portals have become dominant, control personal data, and monetize online presence information as a product. Similar concerns exist in scholarly communication (e.g. \cite{Pooley,Siems,Posada}), with consolidation of services that cover the entire research life-cycle and publishers turning into data analytics companies. Decoupling content hosting from service provision yields a separation between ownership, control, and storage of data on the one hand, and the functionality that can provided on top of it, on the other. It levels the playing field to allow for new entrants and creativity.

The remainder of this paper is structured as follows. Section 2 discusses related work. Our ongoing standards-based specification work is discussed in Section 3. Section 4 reports on an experiment in which data-literature links are pushed towards the repositories that host the resources involved in such links in a manner that complies with the specification. Section 5 provides a summary and an outlook for future work.


\section{Related Work}
\label{Related Work}

With dokie.li \cite{DOKIELI}, Capadisli introduced the use of Linked Data Notifications \cite{LDN} (LDN) with ActivityStreams2 \cite{ACTIVITYSTREAMS} (AS2) payloads as a push-oriented interoperability glue for communication among researchers that publish research outputs in personal data pods and with services that add value to these outputs. His approach provided inspiration for the Researcher Pod \cite{MELLON} and COAR Notify \cite{COARNOTIFY} projects.  

The Researcher Pod project, funded by the Andrew W. Mellon Foundation, explores a scholarly communication system based on the Decentralized Web vision of Tim Berners-Lee \cite{TBL} as was envisioned by Capadisli \cite{CARVEN} and Van de Sompel \cite{HVDSYT}. A Researcher Pod network is highly decentralized: each researcher publishes research outputs in a personal data pod \cite{Solid1,Solid2} that is hosted in a personal web domain. Value is added to these research outputs through interaction with appropriate decoupled services and/or peers. The Researcher Pod network uses LDN and AS2 as core interoperability ingredients, but the AS2 payloads are strictly profiled in order to avoid ambiguous interpretation of notifications. Our \enquote{Event Notifications in Value-Adding Networks} specification is one of the outcomes of this project.

In COAR Notify, the network consists of repositories and providers of overlay review services. A core goal of the project is establishing bidirectional links between preprints and certifications thereof (e.g., stand-alone reviews, endorsements, peer-reviewed articles) as soon as the latter get published. Our specification work is closely aligned with COAR Notify to which we actively contribute, uses the same interoperability ingredients, but has a broader scope regarding what constitutes added value and in which communities the specification can be used. 

ScholeXplorer \cite{SCHOLEXPLORER}, based on the Scholix framework \cite{SCHOLIX,SCHOLIX2}, is a centralized approach to exchange life-cycle information about research outputs, focused on the discovery of research data-literature links. Scholix was partly motivated by the lack of interoperability among various approaches for the point-to-point exchange of data-literature links that had resulted from numerous bilateral agreements \cite{Cross-Linking}. It introduced an approach in which a limited number of hubs, including Crossref, DataCite, and OpenAIRE collect and mine linkage information from a variety of scholarly and web corpora. The hubs support retrieving the link information, which is uniformly expressed according to a link information package schema \cite{SCHOLIX_SCHEMA}; no specific protocol for transporting these packages is selected. ScholeXplorer collects the link information from the hubs and makes it available by means of a query-oriented API allowing, for example, a repository to recurrently poll ScholeXplorer in search of linkages in which its resources are involved. Our specification details an approach that allows expressing and exchanging data-literature (and other) links in a push-oriented, point-to-point manner that leverages W3C standards. Conceptually, it aims to provide the capability to broadcast linkage information directly between the systems that host the artifacts as soon as a link is established.  This in contrast with the Scholix framework that needs to wait for scholarly artifacts to be published, harvested and released in batches, months after the artifact creation date.\footnote{See for example \url{https://scholexplorer.openaire.eu/\#/statistics} for an overview of ScholeXplorer batch frequencies}

\section{The Value-Adding Network}
\label{The Value-Adding Network}

\subsection{Design and Technology Considerations} 
\label{Design and Technology Considerations}

The Value-Adding Network approach utilizes a \emph{point-to-point} communication style, avoiding the need for centralized hubs that can eventually become too powerful to remain trustworthy and too big to fail. The approach is \emph{push-oriented}, with only the relevant nodes being updated about new information as soon at it becomes available. In the highly distributed Value-Adding Network set-up, a pull-oriented approach would introduce a hardly tractable \enquote{which node to pull} challenge in addition to the well-known \enquote{when to poll} conundrum. Interactions are necessarily \emph{asynchronous} owing to the fact that the time between requesting a service and providing a service is unpredictable and could range between almost immediate (e.g., provide a trusted timestamp) to months (e.g., provide a peer-review report). 

The communication patterns in the Value-Adding network are \emph{action-oriented}. The patterns express \emph{when} an activity was initiated, acknowledged, or yielded a result (See Section \ref{Network communication patterns}). The internal workflow of \emph{how} the activity was executed is treated as a black box. As such, message payloads contain only core information (in most cases URL identifiers) to convey in which value-adding activities data nodes, service nodes, and artifacts are involved and what the service results are. The assumption is  that further information about the entities involved is available by using their identifiers in auto-discovery mechanisms supported by protocols such as Signposting \cite{Signposting},  WebID \cite{WebID}, Solid \cite{Solid Protocol}, and Hydra \cite{Hydra}. As such, message payloads can be expressed in a \emph{by reference} rather than \emph{by value} style.

Several candidate protocols exist that could meet the point-to-point, push-oriented, and asynchronous design choices, including WebHooks \cite{Webhooks}, Webmention \cite{Webmention}, and Linked Data Notifications \cite{LDN}. Webhooks, while commonly used, is not formally standardized and does not have a natural tie-in with specific serializations and/or vocabularies. Webmention is uni-directional and, as such, can't be used off-the-shelf for a request-response communication style. Moreover, it does not actually allow transmitting messages and rather merely informs the target resource that the sender resource mentions (links to) it. We chose for Linked Data Notifications in combination with the ActivityStreams 2.0 vocabulary, both W3C standards. That combination was successfully used in the dokie.li work, and has proven to work at scale with ActivityPub \cite{ACTIVITYPUB} implementations that provide a communication channel for millions of users on social networks such as Mastodon \cite{Mastodon}.

\vspace{-0.5em}

\subsection{A network of Data Nodes and Service Nodes}
\label{A network of Data Nodes and Service Nodes}

In a scholarly Value-Adding Network, artifacts are research outputs hosted by \texttt{Data Nodes}, while \texttt{Service Nodes} are systems that add value to these research outputs. Both types of node are essentially web servers that host one or more LDN Inboxes where \texttt{Notifications} can be received. An LDN Inbox is an HTTP endpoint associated with a web server to which notifications can be POSTed. Our specification describes how a value-added service can be triggered by sending an appropriate notification to LDN Inbox of a service node. And it describes how the outcome of providing a value-adding service for an artifact, the \texttt{Service Result}, can be communicated to a data node's LDN Inbox by means of a notification. The notification patterns and payload are discussed in Section \ref{Network communication patterns} and Section \ref{Anatomy of a notification message}, respectively. 

The manner in which value-added services are provided is not in scope of our specification. The internal workflows of a service node is considered as a black box yet could be openly exposed on the basis of other specifications. How nodes, or better the agents that operate on behalf of the nodes, act upon receipt of a notification is also out of scope of our specification. For instance, the notifications could be added to a queue for further processing, e.g., to update the metadata of an artifact, or to write life-cycle information in an \texttt{Event Log}~\cite{EVENTLOG} that is made accessible to support the creation of scholarly knowledge graphs that cover artifacts across data nodes. 

\subsection{Network communication patterns}
\label{Network communication patterns}

At this point, our Value-Adding Network specification details two communication patterns: a one-way pattern and a request-response pattern.

\subsubsection{One-way pattern.} This pattern is used when the agent that operates on behalf of a node sends an unsolicited message about an artifact. One-way patterns are used for sending informational messages pertaining to an artifact to another node in the network. These messages can originate from a data node or a service node. Examples of one-way patterns are:
\begin{itemize}
    \item A repository (data node) informs a subject index (service node) about a new artifact.
    \item A linkage service (service node) informs a repository (data node) about a discovered linkage between a hosted data set artifact and a publication.
\end{itemize}

\subsubsection{Request-response pattern.} This pattern allows agents that act on behalf of a data node to ask a service node to apply a value-added service to one of its artifacts. There is no expectation of an immediate response but definitely the expectation of an eventual response. The pattern doesn't assume there is only one response (or request) but introduces a communication thread (similar to email threads) which will be used to relate notifications to each other. An example of a request-response pattern is:

\begin{itemize}
    \item A repository (data node)  would like a peer review for one of the artifacts and therefore contacts a peer review service (service node) to provide this service. The peer review service can first evaluate the request and acknowledge or reject it. When acknowledged, a peer review can be made available at a later date, and a reference to this peer review can be sent back in a notification to the repository.
\end{itemize}

\subsection{Anatomy of a notification message}
\label{Anatomy of a notification message}

Notifications are exchanged in order to inform agents about added-value that is requested or was provided for artifacts. Network agents send and receive the notifications, data and service nodes provide LDN Inboxes to accept incoming messages.  Value-Adding Networks uses a subset of the ActivityStreams 2.0 vocabulary \cite{ACTIVITYSTREAMS} serialized as JSON-LD~\cite{JSON-LD} payloads. The LDN protocol requires all the payloads to be expressed by default as JSON-LD, but other RDF serializations are allowed. For brevity, all examples in this paper use Turtle syntax \cite{TURTLE}. We refer to our specification for further JSON-LD examples.

Every notification has at least one AS2 type that is a subset of the \texttt{as:Activity} types. The subtypes are limited to those that express facts about artifacts and currently exclude types related to typical social activities (such as \texttt{as:Like}, \texttt{as:Follow}). The subtypes used are:

\begin{itemize}
    \item Activities to request or receive notifications about value adding services: \texttt{as:Announce}, \texttt{as:Offer}, \texttt{as:Accept}, \texttt{as:Reject}, and \texttt{as:Undo}
    \item Activities that express CRUD  (\emph{Create},\emph{Update},\emph{Delete}) life-cycle event on data nodes: \texttt{as:Create}, \texttt{as:Update}, and \texttt{as:Remove}.
\end{itemize}

In addition to the activity types listed above, an activity may have additional subtypes that are specific to an application domain. For instance, when a service node would like to use \texttt{as:Announce} for the endorsement of a publication an extra \texttt{schema:EndorseAction} could be added as activity type. 

Table \ref{table:oneway} shows the mapping between AS2 entities and the Value-Adding Network entities that are used for the notification shown in Listing~\ref{list:as2ex},  which 
illustrates a typical notification that uses the \textbf{one-way pattern} from a service node to a data node. In it, an agent at the end of the service node uses the \texttt{as:Announce} activity type to inform an agent at the end of a data node about the result of adding value to one of the latter's hosted artifacts. In human language, the notification expresses the following:
\say{Fairfield Archive (service node) announces to Springfield Library (data node) a service result concerning Springfield's artifact \url{https://springfield.library.net/artifact/13-02.html.} For this artifact, the \url{https://fairfield.org/archive/version/317831-13210} is a memento.} 

\vspace{-2em}

\begin{table}
\caption{One-way AS2 pattern for notification messages from a service node to a data node.}
\label{table:oneway}
\begin{tabular}{|l|l|}
\hline
AS2 element & Description\\
\hline
\texttt{id} &  Message identifier \\
\texttt{type} & Activity type \\
\texttt{as:actor} & Agent that performed the activity \\
\texttt{as:origin} & Agent responsible for sending the notification \\
\texttt{as:context} & The artifact on the data node for which an value-added service was provided \\
\texttt{as:object} & The result of the value-added service provided for an artifact on the data node \\
\texttt{as:target} & The agent at the data node that is the addressee of the notification \\
\hline
\end{tabular}
\end{table}
\vspace{-1em}

\begin{lstlisting}[label=list:as2ex, caption=A value-adding Activity Streams 2.0 notification using Turtle syntax]
@prefix as: <https://www.w3.org/ns/activitystreams#> .
@prefix ldp: <http://www.w3.org/ns/ldp#> .

<urn:uuid:239FD510-03F4-4B56-B3A0-0D3B92F3826D> a as:Announce ;
  as:actor   <https://fairfield.org/about#us> ;
  as:origin  <https://fairfield.org/system> ;
  as:context <https://springfield.library.net/artifact/13-02.html> ;
  as:object  <urn:uuid:CF21A499-1BDD-4B59-984A-FC94CF6FBA86> ;
  as:target  <https://springfield.library.net/about#us> .

<https://fairfield.org/about#us> a as:Organization ;
  ldp:inbox <https://fairfield.org/inbox> ;
  as:name "Fairfield Archive" .

<https://fairfield.org/system> a as:Service;
  as:name "Fairfield Archive System".

<urn:uuid:CF21A499-1BDD-4B59-984A-FC94CF6FBA86> a as:Relationship ;
  as:object <https://fairfield.org/archive/version/317831-13210> ;
  as:relationship <https://www.iana.org/memento> ;
  as:subject <https://springfield.library.net/artifact/13-02.html> .

<https://springfield.library.net/about#us> a as:Organization;
  ldp:inbox <https://springfield.library.net/inbox/> ;
  as:name "Springfield Library" .
\end{lstlisting}
\vspace{-1em}

\section{Experimental Investigation}

\begin{figure}[h]
\centering
\includegraphics[scale=0.3]{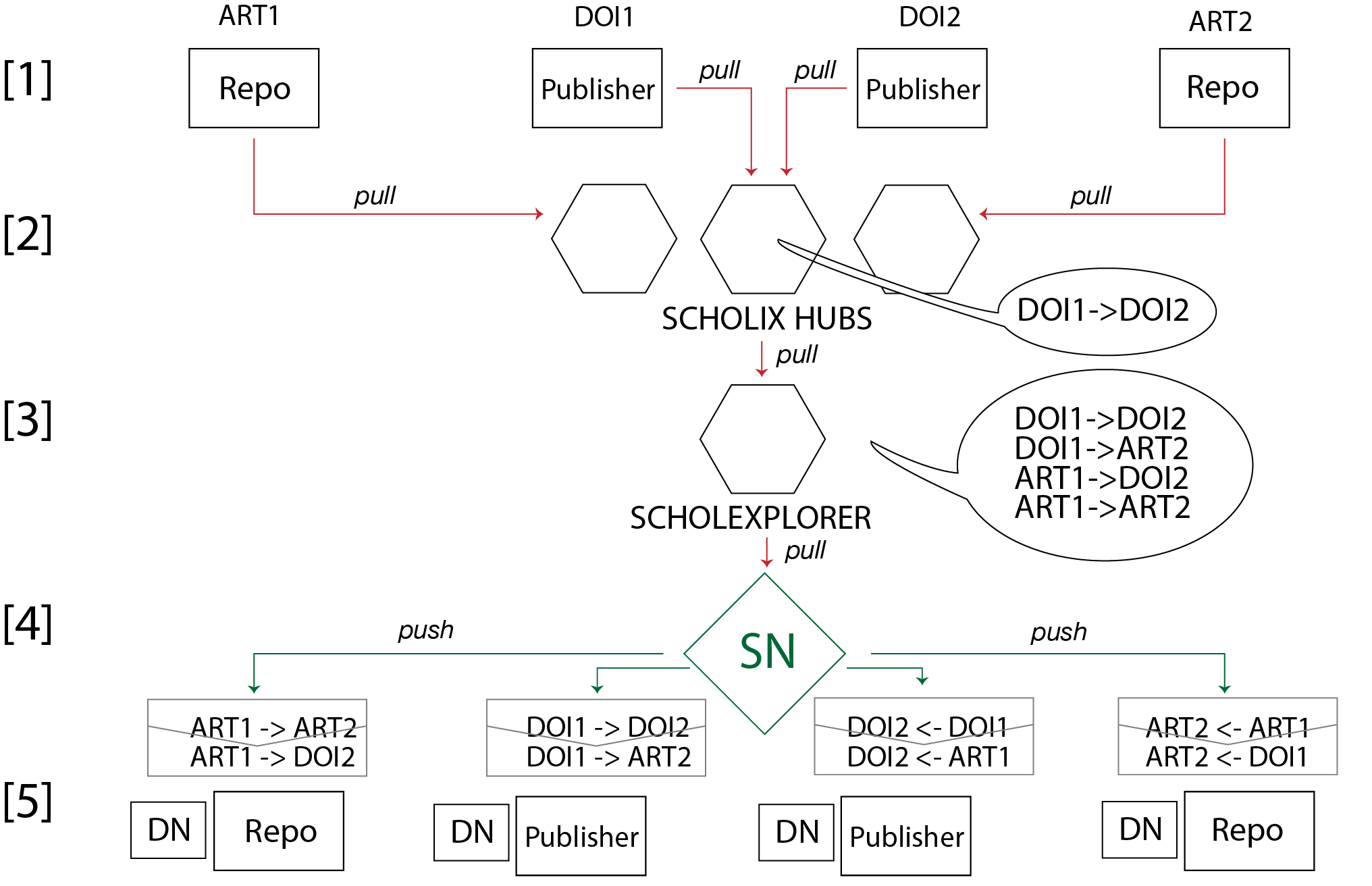}
\caption{Scholix network with an added national service node.}
\label{fig:scholix-network}
\end{figure}

The COAR Notify project has already illustrated the feasibility of real-world use of profiled LDN/AS2 notifications in a request-response pattern between repositories and certification services. We set up an experiment to investigate the one-way pattern where a service node informs a network of data nodes about the \emph{linkages} their artifacts are involved in. We leverage the existing Scholix link base which provides an extensive collection of such linkage information.

In the Scholix network, data-literature linkages are provided via a limited number of hubs such as DataCite, CrossRef and OpenAIRE ([2] in Fig. \ref{fig:scholix-network}). that obtain information from systems that host research outputs ([1] in Fig. \ref{fig:scholix-network}). ScholeXplorer ([3] in Fig. \ref{fig:scholix-network}) is an aggregation service that collects link information from the hubs and makes the result linkages available as a dataset at an irregular frequency. ScholeXplorer can be polled to discover data-literature links between research outputs hosted in the systems covered by the hubs.  For our experiment, we prototype a national service NatServ ([4] in Fig. \ref{fig:scholix-network}) that queries ScholeXplorer for data-literature links in which artifacts hosted by Belgian institutional repositories are involved, and that pushes the link information to the relevant repository according to the Value-Adding Network approach. Using the terminology of that specification, NatServ is a service node that pushes link information to data nodes associated with the institutional repositories ([5] in \ref{fig:scholix-network}). Figure \ref{fig:scholix-network} also shows the research outputs identified by \texttt{ART1}, \texttt{DOI1}, \texttt{DOI2}, \texttt{ART2} hosted by a variety of repositories. The hubs pull information about these artifacts, and as a result of data processing, determine that \texttt{DOI1} is related to \texttt{DOI2}. ScholeXplorer pulls such linkage information from all hubs and further discovers that \texttt{DOI1} is a manifestation of \texttt{ART1} and \texttt{DOI2} is a manifestation of \texttt{ART2}. Finally, our prototype NatServ (SN) pulls data-literature links pertaining to selected repositories from ScholeXplorer and pushes the information using profiled LDN/AS2 notifications to the LDN Inboxes of data nodes associated with the pertinent repositories, e.g. the data node at the left bottom receives information that its ART1 is related to ART2 and DOI2.

\subsection{Experiment}

We downloaded Scholix data for the following Belgian repositories: \emph{Institutional Repository Universiteit Antwerpen} (Antwerpen) , \emph{Ghent University Academic Bibliography} (Biblio) and \emph{Open Repository and Bibliography - University of Liège} (Orbi). These downloads resulted in 711 records for Antwerpen, 1056 for Biblio, and 669 for Orbi. Between May 10 and May 25 2022, we also tried to download Scholix data for the KU Leuven \emph{Lirias} repository and the Vrije Universiteit Brussel \emph{Vrije Universiteit Brussel Research Portal} repository. Although hundreds of Scholix records are available that indicate linkages in which their research outputs are involved, none contained actual URLs of those artifacts.\footnote{see e.g. \url{https://gist.github.com/phochste/4d6ed7f7748c5d74417ac45d38480540}} Since these URLs are needed for the remainder of the experiment, we excluded the Scholix records for these repositories. At the time of writing, our communication with ScholeXplorer representatives have not led to an explanation for this problem.

\subsubsection{Discovery of LDN Inboxes}

With NatServ in possession of the linkages in which artifacts of the three repositories are involved, its next step is to discover the data nodes and associated LDN Inboxes to which linkage information needs to be pushed. The goal is to push the information to the systems that host source as well as target of the links. Since we assume auto-discovery mechanisms for the discovery of pertinent Inboxes, this requires dereferencing the URLs of the all artifacts (expressed in Scholix as PID-URLs, e.g. \texttt{DOI}\footnote{Digital Object Identifier https://doi.org}, \texttt{HANDLE}\footnote{Handle identifier http://hdl.handle.net} , \texttt{PMID}\footnote{PubMed Identifier https://pubmed.ncbi.nlm.nih.gov}, \texttt{PMC}\footnote{PubMed Central https://www.ncbi.nlm.nih.gov/pmc/}, and \texttt{ARXIV}\footnote{arXiv Identifier https://arxiv.org}) up to their landing pages and extracting the link with \texttt{http://www.w3.org/ns/ldp\#inbox} Link relation from its HTTP header.\footnote{See https://www.w3.org/TR/ldn/\#discovery} The URL derferencing process is less trivial than it may seem \cite{KleinBalakireva} and we employed the simplest possible algorithm by following for each link identifier all the HTTP redirects until a page with a HTTP 200 response was found. Obviously, no LDN Inboxes were found because none of the systems involved in linkages currently support Value-Adding Networks, but the excercise remains important as a means to understand the effort that would be involved if they did. 

Table \ref{tab:data-lit-per-o} provides an overview of the complete resolved data-literature link network for each Belgian institution: this is the network of all data-literature links involving artifacts hosted by the institutional repositories. For instance, if one institutional artifact references ten datasets, then this is a network of 10+1 artifact URLs.

\begin{table}

\caption{Number of artifact URLs resolved for the data-literature network of each Belgian institution and time required to resolve PID-URLs to their landing page.}
\label{tab:data-lit-per-o}
\begin{tabular}{|l|r|r|r|r|}
\hline
Scholix Link Provider & \#Records & \# Artifact URLs & \#Resolve time (sec) & time/req\\
\hline
Antwerpen & 711 & 4335 & 695 & 0.978 $\pm$ 0.01 s  \\
Biblio & 1056 & 7189 & 3651 & 3.457 $\pm$ 0.02 s  \\
Orbi & 669 & 3375 & 367 & 0.549 $\pm$ 0.02 s \\
\hline
\end{tabular}

\end{table}

To proceed with the experiment, we created LDN Inboxes for all systems that host artifacts involved in the linkages, using a Solid CSS server as proxy of LDN Inboxes.\footnote{https://github.com/CommunitySolidServer/CommunitySolidServer} We created the inboxes at addresses generated as follows \texttt{https://\{data.host\}/inbox} where \texttt{data.host} is the base url of the landing page. For example, the LDN Inbox URL: \texttt{https://arxiv.org/inbox} will be hosted at: \texttt{http://solidpod-baseurl/arxiv.org/inbox} where \\
\texttt{solidpod-baseurl} is the base URL of the Solid CSS server. 

With Solid CSS, the implementation of LDN Inboxes doesn't require any fundamental changes in institutional repositories to add read-write capabilities to an existing network. In our experiment effortless created hundreds of LDN Inboxes given the fact that: 1) LDN is a subset of the Solid protocol stack \cite{Solid1}, and, 2) an LDN Inbox can be put on any node as long at it can be discovered from the artifact landing page.

In total, 197 [Antwerpen], 245 [Biblio], and 193 [Orbi] LDN Inboxes were created to cover the data-literature network of the institutions.

\subsubsection{Generating notification messages}
Next, we use the guidelines from our specification for sending data about a linking event to transform a ScholeXplorer data-literature link into a profiled LDN/AS2 payload.\footnote{https://www.eventnotifications.net/\#pattern-5-a-service-node-notifies-a-data-node-about-linking-event} The generated payloads are nimble (see snippet in Listing \ref{list:scholix-not}) because we assume that further information pertaining to entities involved can be obtained via auto-discovery mechanisms.  The ActivityStreams 2.0 Activity \texttt{as:Relationship} is used to express a relationship between two artifacts and the mapping that was used is shown in Table \ref{table:scholix}.\footnote{The full Turtle representation of the notification message is available online \url{https://gist.github.com/phochste/afdc7bc5d6a09cec2b12c60755ef1c82}} 

\begin{table}
\caption{One-way AS2 mappings for notification messages in the Scholix experiment.}
\centering
\label{table:scholix}
\begin{tabular}{|l|l|}
\hline
AS2 element & Description\\
\hline
\texttt{as:actor} & We minted the WebID \texttt{https://scholexplorer.openaire.eu/\#about}. \\
\texttt{as:origin} & We minted the WebID \\
 & \texttt{https://mellonscholarlycommunication.github.io/about\#us} \\
\texttt{as:context} & Points to the artifact URL on the target data node the data-literature link \\
 & is about. \\
\texttt{as:object} &  Contains the description data-literature link. An example such a \\
 & data-literature link description is shown in Listing \ref{list:scholix-not}. \\
\texttt{as:target} & This contains the generated (or discovered) LDN Inbox for each \\
 & data-literature link.  \\
\hline
\end{tabular}
\end{table}

\vspace{+1em}

\begin{lstlisting}[escapeinside={(*}{*)}, numbers=left, label=list:scholix-not, caption=Snippet of the \texttt{as:object} of data-literature notification.]
<urn:uuid:240c0091-b271-4e44-87f7-5598da5b24ad>
  a as:Relationship ;
  as:object <https://dx.doi.org/10.5061/dryad.10hq7> ;
  as:relationship <http://www.scholix.org/References> ;
  as:subject <https://biblio.ugent.be/publication/8646849> .
\end{lstlisting}

\subsubsection{Sending notifications to LDN Inboxes}
For each institution, we sent the generated notifications to the simulated LDN Inboxes of all systems for which artifacts are involved in links either as source and target. This doubles the number of messages that is provided in Table \ref{tab:data-lit-per-o}. For instance, if one artifact $A$ references two data sets  $B$ and $C$, then we send two messages for $A$: $A$ references $B$, $A$ references $C$. To $B$ we can send: $A$ references $B$. To $C$ we can send $A$ references $C$. In total four messages.

Table \ref{tab:resolve-url} shows the number of messages that were posted and the typical throughput that is not dependent on the payload of the message.

The generated inboxes and the notifications that were sent to them can be downloaded at Zenodo.\footnote{ \url{https://zenodo.org/record/6555821}} The software to download and transform Scholix into AS2 notifications and to publish notifications to LDN Inboxes is available at GitHub.\footnote{https://github.com/MellonScholarlyCommunication/scholix-client} All experiments were run on a 2 CPU, 1 GB, Linux host connected to the Belnet research network.

\begin{table}
\caption{Sending LDN Notifications for the complete network of three Belgian institutions. The mean posting time for these networks have a constant rate of about 80 notifications per second.}
\label{tab:resolve-url}
\begin{tabular}{|l|l|l|}
\hline
Scholix Link Provider & \# Sent Notifications  & \#Post time (sec) \& time/req\\
\hline
Antwerpen & 8670 & 108s , 80 req/sec \\
Biblio & 14378 & 183s , 78 req/sec \\
Orbi & 6720 & 86s , 78 req/sec \\
\hline
\end{tabular}
\end{table}

\section{Summary and Outlook}
\label{Summary and Outlook}

Our work demonstrated how the introduction of decentralized read-write technology on top of existing research networks provides a push-based mechanism to provide up-to-date information of life-cycle events pertaining to artifacts that are stored in the network. In our case, the network was applied to the distribution of data-literature links by an experimental national NatSev service node (built on top of ScholeXplorer) to data nodes in the network. These experiments demonstrate how the service node was able to inform a network of institutional repositories about data-literature links using push-based LDN Notifications. 

The scalability of the distribution of data-literature links in the Scholix experimental network is dependent on the time it takes to resolve PID-URLs to the landing pages of the artifact. Even with our naive resolve algorithm, which doesn't employ any optimization strategies, the complete Belgian national data-literature network for all known linkages could be distributed within two hours on a small Linux host. 

Our Scholix experiment was  dependent on data mining capabilities of centralized Scholix hubs. The real goal of a distributed Value-Added Network is for linkages to be mined (or cataloged), and sent from the repository that hosts an artifact that is the source of a link to a repository that hosts the artifact that is the target of that link. Research in our Researcher Pod project has already been started to explore how data-mining capabilities can be added to Value-Added networks via orchestration \cite{Orchestrator}.

With our small sample of Scholix providers, we could already create for three Belgian institutions a network with $197+245+193$  data nodes. It would be a bright future for the decentralisation of scholarly communication when all linkage information could be made available in real-time (instead of months after a long publication and review process) and point-to-point instead of requiring a few massive central hubs and gateways. The question how this can be accomplished in a repository landscape that has hardly invested in interoperability affordances since the broad adoption of OAI-PMH \cite{OAIPMH} in the early 2000s. 

Our hope is that COAR’s \enquote{Next Generation Repositories} vision will get a broader uptake. Important steps are taken in the COAR Notify project with the first concrete implementations that make profiled LDN/AS2 notifications a reality for communication between repositories and review services. Arcadia has recently recommended a 4 year,  US\$4 million grant \cite{Arcadia} to accelerate COAR Notify enhancing the role of repositories, thereby transforming scholarly communications, making it more research-centric, community-governed, and responsive to the diverse needs across the globe.

\subsubsection{Acknowledgements} 
This work is funded by the Andrew W. Mellon Foundation (grant number: 1903-06675) and supported by SolidLab Vlaanderen (Flemish Government, EWI and RRF project VV023/10). The authors would like to thank the COAR Notify (Kathleen Shearer, Paul Walk, Martin Klein) for involving them from the outset, allowing them to provide input and gain valuable insights for their generic specification effort.

%
%
%
\bibliographystyle{splncs04}

\begin{thebibliography}{8}

\bibitem{Rethinking}
Van de Sompel, H., Payette, S., Erickson, J., Lagoze, C.,  Warner, S. : Rethinking Scholarly Communication. D-Lib Magazine, 10(9) (2004)
\url{https://www.dlib.org/dlib/september04/vandesompel/09vandesompel.html}

\bibitem{ROOSENDAAL}
Roosendaal, H., Geurt, P.: Forces and functions in scientific communication:
an analysis of their interplay. (1998) \url{https://perma.cc/5HYM-BEKF}. Captured 3 June 2019

\bibitem{CARVEN}
Capadisli, S.: Linked Research on the Decentralised Web. (2020) \url{https://csarven.ca/linked-research-decentralised-web}

\bibitem{NextGenRep}
Rodrigues, E., Bollini, A., Cabezas, A., Castelli, D., Carr, L., Chan, L., Humphrey, C., Johnson, R., Knoth, P., Manghi, P., Matizirofa, L., Perakakis, P., Schirrwagen, J., Selematsela, D., Shearer, K., Walk, P., Wilcox, D., Yamaji, K.: Next Generation Repositories: Behaviours and Technical Recommendations of the COAR Next Generation Repositories Working Group. Zenodo. (2017)  \url{https://doi.org/10.5281/zenodo.1215014}

\bibitem{paper_link_preprints_reviewed}
Cabanac, G., Oikonomidi, T. \& Boutron, I. Day-to-day discovery of preprint–publication links. Scientometrics 126, 5285–5304 (2021). \url{https://doi.org/10.1007/s11192-021-03900-7}

\bibitem{MellonSpec}
Hochstenbach, P., Vander Sande, M., Dedecker, R., Walk, P., Klein, M., Van de Sompel, H. : Event Notifications in Value-Adding Networks. (2020). \url{https://www.eventnotifications.net} Accessed 14 Jul 2022

\bibitem{socially_aware_cloud_storage}
Berners-Lee, T : Socially Aware Cloud Storage. (2009) \url{https://www.w3.org/DesignIssues/CloudStorage.html}

\bibitem{Pooley}
Pooley, J. : Surveillance Publishing. Elephant in the Lab. (2022) \url{https://doi.org/10.5281/zenodo.6384605}

\bibitem{Siems}
Siems, R. : Das Lesen der Anderen: Die Auswirkungen von User Tracking auf Bibliotheken. O-Bib. Das Offene Bibliotheksjournal / Herausgeber VDB, 9(1), 1–25. (2022) \url{https://doi.org/10.5282/o-bib/5797}

\bibitem{Posada}
Posada, A., Chen, G.: Inequality in Knowledge Production: The Integration of Academic Infrastructure by Big Publishers. ELPUB 2018, Jun 2018, Toronto, Canada. \url{dx.doi.org/10.4000/proceedings.elpub.2018.30}

\bibitem{DOKIELI}
Capasidli, S.: Dokieli \url{https://dokie.li}

\bibitem{LDN}
Capadisli, S., Guy, A.: Linked Data Notifications. W3C Recommendation. (2017) 
\url{https://www.w3.org/TR/ldn/}

\bibitem{ACTIVITYSTREAMS}
Snell, J., Prodromou, E.: Activity Streams 2.0. W3C Recommendation. (2017) \url{https://www.w3.org/TR/activitystreams-core/}

\bibitem{MELLON}
"Scholarly Communications in the Decentralized Web".
\url{https://mellon.org/grants/grants-database/grants/ghent-university/1903-06675/}. Accessed 20 May 2022

\bibitem{COARNOTIFY}
"The Notify Project". \url{https://www.coar-repositories.org/notify/}. Accessed 20 May 2022

\bibitem{TBL}
Berners-Lee, T.: \url{https://www.w3.org/DesignIssues/Principles.html}. 
\bibitem{HVDSYT}
Van de Sompel, H.: Scholarly Communication: Deconstruct \& Decentralize? \url{https://www.youtube.com/watch?v=o4nUe-6Ln-8}

\bibitem{Solid1}
Sambra, A.V., Mansour, E., Hawke, S., Zereba, M., Greco, N., Ghanem, A., Zagidulin, D., Aboulnaga, A., Berners-Lee, T.: Solid: a platform for decentralized social applications based on linked data. Tech. rep., Technical report, MIT CSAIL
\& Qatar Computing Research Institute (2016)
\url{http://emansour.com/research/lusail/solid\_protocols.pdf}

\bibitem{Solid2}
Mansour, E., Sambra, A.V., Hawke, S., Zereba, M., Capadisli, S., Ghanem, A.,
Aboulnaga, A., Berners-Lee, T.: A demonstration of the solid platform for social web applications. In: Proceedings of the 25th International Conference Companion on World Wide Web. pp. 223–226 (2016) \url{https://dl.acm.org/doi/10.1145/2872518.2890529}

\bibitem{SCHOLEXPLORER}
La Bruzzo, S., Manghi, P. : OpenAIRE ScholeXplorer Service: Scholix JSON Dump (5.0) [Data set]. Zenodo. (2022) \url{https://doi.org/10.5281/zenodo.6338616}

\bibitem{SCHOLIX}
Burton, A., Koers, H., Manghi, P., Stocker, M., Fenner, M., Aryani, A., La Bruzzo, S., Diepenbroek, M. and Schindler, U. : The Scholix framework for interoperability in data-literature information exchange. D-Lib Magazine, 23(1/2) (2017) \url{http://www.dlib.org/dlib/january17/burton/01burton.html}

\bibitem{SCHOLIX2}
Burton, A., Koers, H., Manghi, P., La Bruzzo. S., Aryani, A., Diepenbroek, M., Schindler, U. :  On Bridging Data Centers and Publishers: The Data-Literature Interlinking Service. In: Garoufallou, E., Hartley, R., Gaitanou, P. (eds) Metadata and Semantics Research. MTSR 2015. Communications in Computer and Information Science, vol 544. Springer, Cham. (2015) \url{https://dx.doi.org/10.1108/PROG-06-2016-0048}

\bibitem{Cross-Linking}
Callaghan, S., Tedds, J., Lawrence, R., Murphy, F., Roberts, T., and Wilcox, W. : Cross-Linking Between Journal Publications and Data Repositories: A Selection of Examples. International Journal of Digital Curation. (2014) \url{https://doi.org/10.2218/ijdc.v9i1.310}

\bibitem{SCHOLIX_SCHEMA}
"Scholix metadata schema for exchange of scholarly links". \url{http://www.scholix.org/schema} Accessed 24 May 2022

\bibitem{Signposting}
"Signposting the Scholarly Web". \url{https://signposting.org}. Accessed at 20 May 2022 

\bibitem{WebID}
W3C. "WebID". \url{https://www.w3.org/wiki/WebID}. Accessed at 25 May 2022

\bibitem{Solid Protocol}
Capadisli, S., Berners-Lee, T., Verborgh, R., Kjensmo, K. : Solid Protocol Version 0.9.0, 2021-12-17. \url{https://solidproject.org/TR/protocol} Accessed at 25 May 2022

\bibitem{Hydra}
Lanthaler, M.: Hydra Core Vocabulary. W3C Unofficial Draft (2021) \url{https://www.hydra-cg.com/spec/latest/core/}

\bibitem{Webhooks}
"Webhooks – The Definitive Guide". (2022) \url{https://webhook.net}. Accessed at 20 May 2022

\bibitem{Webmention}
Parecki, A. : Webmention . W3C Recommendation. (2017) \url{https://www.w3.org/TR/webmention/}

\bibitem{ACTIVITYPUB}
Lemmer-Webber, C., Tallon, J., Shepherd, E., Guy, A., Prodromou, E.: ActivityPub. W3C Recommendation. (2019) \url{https://www.w3.org/TR/activitypub/}

\bibitem{Mastodon}
"Giving social networking back to you - Mastodon". \url{https://joinmastodon.org}. Accessed at 20 May 2022

\bibitem{EVENTLOG}
Vander Sande, M., Hochstenbach, P., Dedecker, R., Werbrouck, J. : Artefact Lifecycle Event Log. (2022) \url{https://mellonscholarlycommunication.github.io/spec-eventlog/}

\bibitem{JSON-LD}
Sporny, M., Longley, D., Kellogg, G., Lanthaler, M., Champin, P.A., Lindström, N. :  JSON-LD 1.0. W3C Recommendation. (2014) \url{https://www.w3.org/TR/json-ld/}

\bibitem{TURTLE}
Beckett, D., Berners-Lee, T., Prud'hommeaux, E., Carothers, G. : RDF 1.1 Turtle". W3C Recommendation. (2014) \url{https://www.w3.org/TR/turtle/}

\bibitem{KleinBalakireva}
Klein, M., Balakireva, L.: On the Persistence of Persistent Identifiers of the Scholarly Web. In: Hall, M., Merčun, T., Risse, T., Duchateau, F. (eds) Digital Libraries for Open Knowledge. TPDL 2020. Lecture Notes in Computer Science(), vol 12246. Springer, Cham. (2020) \url{https://doi.org/10.1007/978-3-030-54956-5\_8}

\bibitem{ZenodoData}
Hochstenbach, Patrick : Notifications in Value-Adding Networks (1.0.0). Zenodo.  (2022). \url{https://doi.org/10.5281/zenodo.6555821}

\bibitem{ScholixClient}
Hochstenbach, P.: Mellon Scholary Communication, (2022), GitHub repository. \url{https://github.com/MellonScholarlyCommunication/scholix-client}

\bibitem{Orchestrator}
Vander Sande, M., Hochstenbach, P., Dedecker, R., Werbrouck, J.: Orchestrator for a decentralized Web network. Working Draft, 26 November 2021
 (2021) \url{https://mellonscholarlycommunication.github.io/spec-orchestrator/}
 
\bibitem{OAIPMH}
Lagoze, C., Van de Sompel H., Nelson, M., Warner, S. : The Open Archives Initiative Protocol for Metadata Harvesting. Protocol Version 2.0 of 2002-06-14 (2015) \url{http://www.openarchives.org/OAI/openarchivesprotocol.html} 

\bibitem{Arcadia}
COAR. "COAR welcomes significant funding for the Notify Project". \url{https://www.coar-repositories.org/news-updates/coar-welcomes-significant-funding-for-the-notify-project/}. Accessed 25 May 2022

\end{thebibliography}
%

\end{document}